\documentclass[%
 reprint,
 superscriptaddress,
amsmath,amssymb,
aps,
pra,
]{revtex4-2}
\usepackage{lipsum}
\usepackage{color}
\usepackage{marvosym}
\usepackage{graphicx}
\usepackage{dcolumn}
\usepackage{bm}
\usepackage{appendix}
\usepackage{orcidlink}
\hypersetup{colorlinks=true, linkcolor=blue, citecolor=blue, urlcolor=blue}

\begin{document}

\title{Localization and scattering of a photon in quasiperiodic qubit arrays}

\author{Xinyin Zhang~\orcidlink{0009-0007-9108-0534}}
\affiliation{Laboratory of Quantum Engineering and Quantum Metrology, School of Physics and Astronomy, {Sun Yat-Sen University} (Zhuhai Campus), Zhuhai 519082, China}
\affiliation{Institute of Quantum Precision Measurement, State Key Laboratory of Radio Frequency Heterogeneous Integration, College of Physics and Optoelectronic Engineering, {Shenzhen University}, Shenzhen 518060, China}

\author{Yongguan Ke~\orcidlink{0000-0003-2740-0007}} 
\altaffiliation{Email: keyg@szu.edu.cn}
\affiliation{Institute of Quantum Precision Measurement, State Key Laboratory of Radio Frequency Heterogeneous Integration, College of Physics and Optoelectronic Engineering, {Shenzhen University}, Shenzhen 518060, China}

\author{Zhengzhi Peng~\orcidlink{0009-0002-8156-4758}}
\affiliation{Laboratory of Quantum Engineering and Quantum Metrology, School of Physics and Astronomy, {Sun Yat-Sen University} (Zhuhai Campus), Zhuhai 519082, China}
\affiliation{Institute of Quantum Precision Measurement, State Key Laboratory of Radio Frequency Heterogeneous Integration, College of Physics and Optoelectronic Engineering, {Shenzhen University}, Shenzhen 518060, China}

\author{Zuorui Chen}
\affiliation{Department of Mechanical and Aerospace Engineering, {The Hong Kong University of Science and Technology}, Clear Water Bay, Hong Kong, China}

\author{Wenjie Liu~\orcidlink{0000-0002-7522-9526}}
\affiliation{School of General Education, Dalian University of Technology, Panjin 124221, China}

\author{Li Zhang~\orcidlink{0000-0001-5486-2691}}
\affiliation{Institute of Quantum Precision Measurement, State Key Laboratory of Radio Frequency Heterogeneous Integration, College of Physics and Optoelectronic Engineering, {Shenzhen University}, Shenzhen 518060, China}

\author{Chaohong Lee~\orcidlink{0000-0001-9883-5900}}
\affiliation{Institute of Quantum Precision Measurement, State Key Laboratory of Radio Frequency Heterogeneous Integration, College of Physics and Optoelectronic Engineering, {Shenzhen University}, Shenzhen 518060, China}
\affiliation{{Quantum Science Center of Guangdong-Hong Kong-Macao Greater Bay Area} (Guangdong), Shenzhen 518045, China}

\date{\today}

% =========================================Abstract=======================================

\begin{abstract}
We study the localization and scattering of a single photon in a waveguide coupled to qubit arrays with quasiperiodic spacings. 
As the quasiperiodic strength increases, localized subradiant states with extremely long lifetime appear around the resonant frequency and form a continuum band. 
In stark contrast to the fully disordered waveguide QED where all states are localized, we analytically find that the fraction of localized states is up to $(3-\sqrt{5})/2$ when the modulation frequency is $(1+\sqrt{5})/2$.  
The localized and delocalized states can be related to excitation in flat and curved inverse energy bands under the approximation of large-period modulation.
When the quasiperiodic strength is weak, an extended subradiant state can support the transmission of a photon.
However, as the quasiperiodic strength increases, localized subradiant states can completely block the transmission of a single photon in resonance with the subradiant states, and enhance the overall reflection.
At a fixed quasiperiodic strength, we also find mobility edge in transmission spectrum, below and above which the transmission is either turned on and off as system size increases.
Our work give new insights into the localization in non-Hermitian systems.
\end{abstract}

% ========Main body=========================================================
\maketitle

\section{Introduction}
Photons in a waveguide coupled to natural or artificial atoms, known as the waveguide quantum electrodynamics (WQED) system, serves as an important platform for engineering light-atom interaction and developing practical applications for quantum information processing~\cite{Roy2017,Chang2018,Gutzler2021,RevModPhys.95.015002}.   
WQED can be realized in a variety of experimental systems, involving cold atoms~\cite{Goban2015,corzo2019waveguide,PhysRevLett.128.073601}, superconducting qubits~\cite{van2013photon,liu2017quantum,Kim2021}, quantum dots~\cite{doi:10.1126/science.ade9324} and solid-state defects~\cite{Lodahl2015,Foster2019,Jeannic2021}.  
With current quantum technologies, it is possible to prepare and detect excitation states in atoms facilitated by a waveguide~\cite{corzo2019waveguide,leong2020large}.
Recently, many novel excitation states such as twilight states~\cite{Ke2019}, bound states~\cite{Zhang2020,Ke2020}, self-localized states~\cite{Zhong2020}, and chaos states~\cite{Poshakinskiy2021,Alexander2020} have been predicted in theoretical works.
It is appealing to understand the relation between excitation states and photon scattering~\cite{Brehm2021o}.
Taking an equal-spacing qubit array as an example, subradiant states can enhance transmission~\cite{Brehm2021o} and inelastic scattering of photons~\cite{Ke2019}.
Disorder is ubiquitous in the fabrication of WQED systems, and it can affect both photon scattering and excitation states~\cite{PhysRevX.7.011034,Mirza2017,Mirza:18,RevModPhys.91.021001,Jen_2021,fayard2021many,Song:21,Brehm2021,PhysRevA.105.023717,PhysRevA.106.043723,PhysRevLett.131.253602,PhysRevA.109.013720,PhysRevResearch.6.013159,PhysRevResearch.6.013320,7ywr-kd8y,c4hd-8vfj}.
Disorder at both atomic positions and transition frequencies can suppress the transmission of a single photon in a bidirectional WQED system~\cite{Mirza2017,Song:21,Brehm2021}, indicating Anderson localization. 
It has been reported that all single-excitation states are localized in the thermodynamic limit of a large qubit array even with tiny randomness in positions~\cite{fayard2021many}.
Moreover, disorder also plays a crucial role in multi-photon scattering and photon-photon correlation.
Recent research reveals that disorder in transition frequencies can induce photon antibunching through destructive interference of photon scattering paths~\cite{mldt-d59t}.
However, the relation between localization and scattering of photons is less known in quasiperiodic structure as an intermediate phase between fully periodic and fully disordered media.
Compared to disordered media, quasiperiodic media can also cause localization, but in a different and more controllable way~\cite{PODDUBNY20101871}, which may be more beneficial for practical applications.
A question naturally arises: How do excitation states affect the scattering of a photon in a quasiperiodic WQED system? 

%%%%%%%%%%%%%%%%%%%%%%%%%%%%%
\begin{figure}[!b]
	\includegraphics[width=0.5\textwidth]{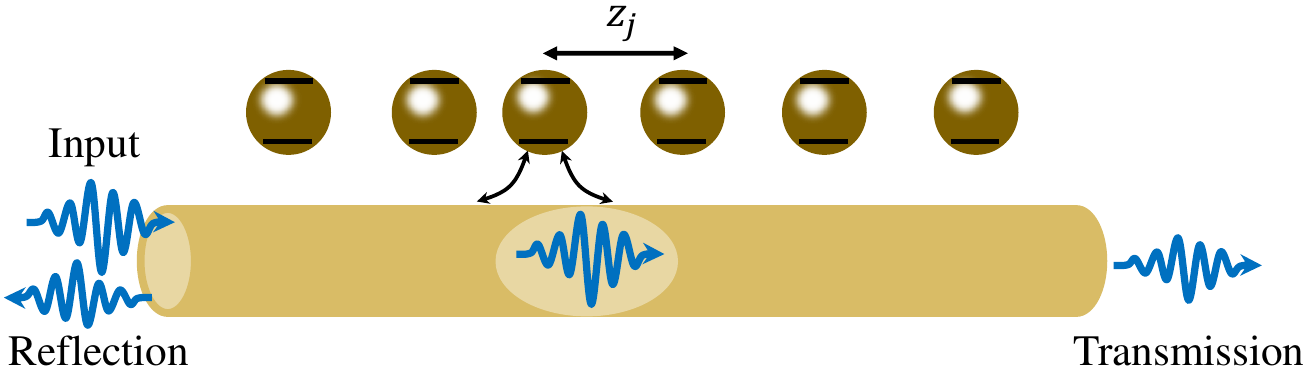}
	\caption{Schematic diagram of a photon in a waveguide scattered by qubit arrays with quasiperiodic spacing. The positions are arranged as $z_j=d[j+\delta\cos (2\pi\beta j +\theta)]$ with averaged spacing $d$, modulation strength $\delta$, modulation frequency $\beta$, and modualtion phase $\theta$.  
	}\label{fig:N}
\end{figure}
%%%%%%%%%%%%%%%%%%%%%%%%%%%%%

Here, we study the relation between localization of an excitation and scattering of a photon in quasiperiodic WQED, as depicted in Fig.~\ref{fig:N}.
The total number of excitations and photons is conserved, and thus we can treat the scattering in the decoupled subspace with a fixed particle number.
As the quasiperiodic strength increases, there appears to be a continuum band of localized excitation states near the transition frequency of the atoms. 
The localized states are subradiant states which have extremely long lifetimes.
Unlike fully disordered case where all states are localized~\cite{fayard2021many}, only a fraction of excitation states are localized states.
Under approximation of the large-period modulation, we find that the delocalized states and localized states come from excitation in flat inverse energy bands and curved inverse energy bands, respectively. 
The fraction of localized states is up to $(3-\sqrt{5})/2$ when the modulation frequency is $(1+\sqrt{5})/2$.  
The subradiant extended states and localized states will support and suppress the transmission of a single photon, respectively.
The overall reflection increases with the quasiperiodic strength due to the stronger localization of excitation. 
Surprisingly, we also find mobility edge in the transmission spectrum that separates localization phase (block of photon) and delocalization phase (transmission of photon).

\section{Model and methods}
\subsection{Quasiperiodic Waveguide QED}
We consider the propagation of photons in a waveguide coupled to two-level atoms with quasiperiodic spacing, which is characterized by the Hamiltonian in real space that contains three parts~\cite{Shen2007}, 
\begin{equation}
	H_{R} = {H_A} + {H_F} + {H_I}.
\end{equation}
Here, $H_A$ describes the energy of excitations,
\begin{equation}
{H_A} = \sum\limits_j^{} {\hbar {\omega _0}b_j^\dag } {b_j},
\end{equation}
where $\omega_0$ is the uniform resonant frequency of the two levels in atoms,
${b_j^{\dag}}$ ($b_j$) is the excitation creation (annihilation) operator on the $j$th atom.
 ${H_F}$ describes the propagating photon in the waveguide along $z$ direction with light velocity $c$,
\begin{equation}
{H_F} = i\hbar c\int_{ - \infty }^{ + \infty } {dz \big[ {a_L^\dag (z)\frac{\partial }{{\partial z}}{a_L}(z) - a_R^\dag (z)\frac{\partial }{{\partial z}}{a_R}(z)} \big]},
\end{equation}
where $a_L^\dag(z)$ ($a_L(z)$) and $a_R^\dag(z)$ ($a_R(z)$) are the creation (annihilation) operators for a photon propagating to the left and right at position $z$, respectively.
${H_I}$ describes the interaction between excitation and photon,
\begin{equation}
	{H_I} = \sum\limits_{j=1}^{N} {\hbar g\int_{}^{} {dz\delta (z - {z_j})\left[ {\big( {a_L^\dag (z) + a_R^\dag (z)} \big){b_j} + h.c.} \right]} },
\end{equation}
where $g$ is the coupling strength, and the total number of atoms is $N$. The position of the $j$th atom is given by
\begin{equation}
z_j=d[j+\delta \cos (2\pi\beta j+\theta)],
\end{equation}
where $d$ is the spacing constant, $\delta$ is the quasi-periodic strength, $\beta$ is the modulation frequency, and $\theta$ is the modulation phase. 
The qubit positions become quasiperiodic when $\beta$ is an irrational number.

Alternatively, by making a Fourier transform, 
\begin{eqnarray}
	a_L(z)&=&\frac{1}{\sqrt{2\pi}}\int_{-\infty}^0 a(k) e^{i kz}dk,\nonumber  \\
	a_R(z)&=&\frac{1}{\sqrt{2\pi}} \int_{0}^{+\infty} a(k) e^{i kz}dk, 
\end{eqnarray}
we can derive the Hamiltonian in the momentum space,
\begin{eqnarray}\label{eq:H0}
	H_{M}&=&\int_{-\infty}^{+\infty} \hbar\omega_{k}a(k)^{\dag}a(k)^{\vphantom{\dag}}dk
	+\sum\limits_{j}\hbar\omega_{0}b_{j}^{\dag}b_{j}^{\vphantom{\dag}} \\
	&+&\frac{\hbar g}{\sqrt{2\pi}}\sum\limits_{j}\int_{-\infty}^{+\infty} 
	(b_{j}^{\dag}a{(k)}e^{i k z_{j}}+b_{j}a{(k)}^{\dag}e^{-i k z_{j}})dk.\nonumber
\end{eqnarray}
Here, $a{(k)}$ is the annihilation operator for a photon with a wave vector $k$.
$\omega_k=c|k|$ is the frequency of a single photon with light velocity $c$. 

\subsection{Transfer-matrix method in real space} \label{Sec41}
We consider the scattering of a photon with momentum $\kappa$ and frequency $\omega_\kappa=c|\kappa|$ propagating rightward in the waveguide.
The transmission and reflection of the photon are strongly modified by the interaction with a quasiperiodic qubit array.
To solve the scattering problem of a single photon, we first introduce the transfer-matrix method in the real space.

We write an ansatz for a general eigenstate~\cite{Zheng2013},
\begin{widetext}
	\begin{equation}
		|{E_k}\rangle  = \sum\limits_j^{} {{e_j}b_j^\dag |0\rangle }  + \int {dz{\phi _L}(z)c_L^\dag (z)|0\rangle }  + \int {dz{\phi _R}(z)c_R^\dag (z)|0\rangle }. 
	\end{equation}
	with
	\begin{eqnarray} \label{Coefficients}
		{\phi _R}(z) &=& \frac{{{e^{i\kappa z}}}}{{\sqrt {2\pi } }}\left[ {\theta ({z_1} - z) + {t_1^2}\theta (z - {z_1})\theta ({z_2} - z) + {t_2^3}\theta (z - {z_2})\theta ({z_3} - z) + ... + {t(\kappa)}\theta (z - {z_N})}, \right] \\ 
		{\phi _L}(z) &=& \frac{{{e^{ - i\kappa z}}}}{{\sqrt {2\pi } }}\left[ {r(\kappa)\theta ({z_1} - z) + {r_{1}^2}\theta (z - {z_1})\theta ({z_2} - z) + {r_{2}^3}\theta (z - {z_2})\theta ({z_3} - z) + ... + {r_{N - 1}^N}\theta (z - {z_{N - 1}})\theta ({z_N} - z)} \right]. \nonumber
	\end{eqnarray}
\end{widetext}
Here, $\theta(z)$ is the Heaviside function.
$r_j^{j+1}$ ($t_j^{j+1}$) is the amplitude of reflection (transmission) between the $j$th and $(j+1)$th atoms, while $r(\kappa)\equiv r_{0}^1$ and $t(\kappa)\equiv t_{N}^{N+1}$ are the amplitudes of reflection by the first atom and transmission through the last atom, respectively. 
This ansatz satisfies the eigenvalue equation $H_R |{E_{\kappa}}\rangle  = {E_{\kappa}}|{E_{\kappa}}\rangle $, with $E_{\kappa}=\hbar c |\kappa|$.
A relation between backward reflection and forward transmission can be obtained using the transfer-matrix method~\cite{PhysRevLett.131.103604,Ke2025},
\begin{eqnarray}
	\left( {\begin{array}{*{20}{c}}
			{{t_\kappa}}\\
			0
	\end{array}} \right) &=& M_N M_{N-1}...M_1 \left( {\begin{array}{*{20}{c}}
			1\\
			{{r_\kappa}}
	\end{array}} \right)\nonumber \\
	&=&\left( {\begin{array}{*{20}{c}}
			{{T_{11}}}&{{T_{12}}}\\
			{{T_{21}}}&{{T_{22}}}
	\end{array}} \right)\left( {\begin{array}{*{20}{c}}
			1\\
			{{r_\kappa}}
	\end{array}} \right),
\end{eqnarray}
with the transfer matrix on the $j$th atom, 
\begin{equation}
	M_j=\left( {\begin{array}{*{20}{c}}
			{  {{\left( {f_\kappa + 1} \right)}^2} - {{f_\kappa}^2}}&{ 2f_\kappa{e^{ - i2\kappa{z_j}}}}\\
			{-2f_\kappa{e^{2i\kappa{z_j}}}}&{-f_\kappa^2 + {{\left( {1 - f_\kappa} \right)}^2}}
	\end{array}} \right).
\end{equation}
Here, $f_\kappa = {{i{\Gamma _0}}}/{{[2({\omega _0} - {c}|\kappa|)]}}$, and $\Gamma_0=g^2/c$ is the decay rate of a single excited qubit.
Finally, the reflection and transmission are respectively given by
\begin{eqnarray}
	{r_\kappa} &=&  - \frac{{{T_{21}}}}{{{T_{22}}}}, \nonumber \\
	{t_\kappa} &=& {T_{11}} - \frac{{{T_{12}}{T_{21}}}}{{{T_{22}}}},
\end{eqnarray}
and they satisfy a conserved relation $|r_\kappa|^2+|t_\kappa|^2=1$.

\subsection{Green function method in momentum space} 
\label{Sec42}
We can alternatively use the Green function method to calculate the reflection and transmission.
After injecting a right-propagating photon with momentum ($\kappa>0$) into the waveguide, it interacts with the emitters and is transferred into an excitation.
The propagation of an excitation in the qubit array is governed by the Green function, which is determined by
\begin{equation}
	{G^{ - 1}}(\omega ) = \omega -H_{eff}, \label{green}
\end{equation}
with the effective Hamiltonian for an excitation~\cite{Ke2019,Ke2020},
\begin{eqnarray}\label{effectiveHam}
	H_{eff}=\sum\limits_{j}\omega_{0}b_{j}^{\dag}b_{j}^{\vphantom{\dag}}
	-i \Gamma_0 \sum\limits_{j,l} b_j^\dag b_l e^{i\omega/c|z_l-z_j|}.
\end{eqnarray}
Here,  $\Gamma_0=g^2/c$ is the radiative decay rate, and the phase constant is defined as $\varphi=\omega d/c$, depending on the frequency of input photon $\omega=c|\kappa|$, spacing constant $d$ and the light velocity $c$.
The hopping of excitation from $z_l$ to $z_j$ points is assisted by the photon emitted from the $l$th qubit and consequently reabsorbed by the $j$th qubit.
When the spacing constant is small enough, the phase constant in the effective Hamiltonian can be replaced by $\varphi=\omega_0 d/c$, which is the so-called Markov approximation~\cite{Ke2019}.
Based on the Green function method~\cite{PhysRevLett.131.103604,Ke2025}, the reflection coefficient is given by
\begin{equation}
	r_{\kappa} =-i\Gamma_0 \sum_{j,j'}G_{j,j'}(\omega_{\kappa})e^{i\omega_{\kappa}/c (z_j+z_{j'})}, \label{reflectionk}
\end{equation}
and the transmission is given by
\begin{equation}
	t_{\kappa} = 1 - i{\Gamma _0}\sum\limits_{j,j'}^{} {{{G_{j,j'}}(\omega_k ) e^{i\omega_k /c({z_{j'}} - {z_j})}}}, \label{transmission}
\end{equation}
and they also satisfy $|r_{\kappa}|^2+|t_{\kappa}|^2=1$.
Because the reflection can be immediately obtained once the transmission is known, we focus on the reflection of a single photon. 
To make it clear, we expand the Green function in terms of the eigenvalues $\{\omega_n\}$ and eigenstates $|\{\psi_n\}\rangle$ of the effective Hamiltonian,
\begin{equation}
	G_{j,j'}(\omega_{\kappa})=\sum\limits_{n}\frac{\psi_n(j)\psi_n(j')}{\omega_{\kappa}-\omega_n}, \label{Green}
\end{equation}
where the eigenstate $|\psi_n\rangle$ has been normalized.
By combining Eqs.~\eqref{Green} and \eqref{reflectionk}, we can obtain the following result 
\begin{equation}
	r_{\kappa} = -i{\Gamma _0}\sum\limits_{j,j',n}^{} {{e^{i\omega_{\kappa} /c({z_{j'}} + {z_j})}}\frac{\psi_n(j)\psi_n(j')}{\omega_{\kappa}-\omega_n}}, \label{transmission1}
\end{equation} 
When the frequency of the photon is in resonance with the energy of an excitation state [i.e., $\omega_{\kappa}=\textrm{Re}(\omega_n)$],
then the denominator only takes the value of the imaginary part of the eigenvalue $\textrm{Im}(\omega_n)$.
If the eigenstate $|\psi_n\rangle$ is a subradiant state with $\textrm{Im}(\omega_n)\ll \Gamma_0$, then reflection is mainly determined by the properties of the $n$th eigenstate.
Thus, reflection may give information about the localization property of subradiant states.

We have to emphasize that both the transfer-matrix method and Green function method give exactly the same results and have their advantages. 
The first method can handle much larger systems, and the second method gives an intimate relation between the scattering and subradiant states. 
In the next section, we will show the reflection of a single photon and its relation to the properties of subradiant eigenstates.

\section{Results}
Our major goal is to explore the localization properties of an excitation and their relation to the scattering of a photon.
According to our previous study~\cite{Ke2019}, the incoherent scattering of photons can be enhanced via subradiant states of excitations.
Before proceeding to the scattering of a single photon, we first study the localization of a single excitation in a quasiperiodic qubit array via the effective Hamiltonian.
We reveal the localization of a single excitation in Sec.~\ref{Sec3}, give the fraction of localization under periodic approximation in Sec.~\ref{Sec5}, show how the localized single-excitation state affects the reflection of a single photon in Sec.~\ref{Sec4}, and finally, discuss the mobility edge in transmission spectrum of a single photon in Sec.~\ref{Sec6}.

\subsection{Continuum band of Localized states} \label{Sec3}
%
%%%%%%%%%%%%%%%%%%%%%%%%%%%%%%
\begin{figure}[!htp]
	\includegraphics[width=0.5\textwidth]{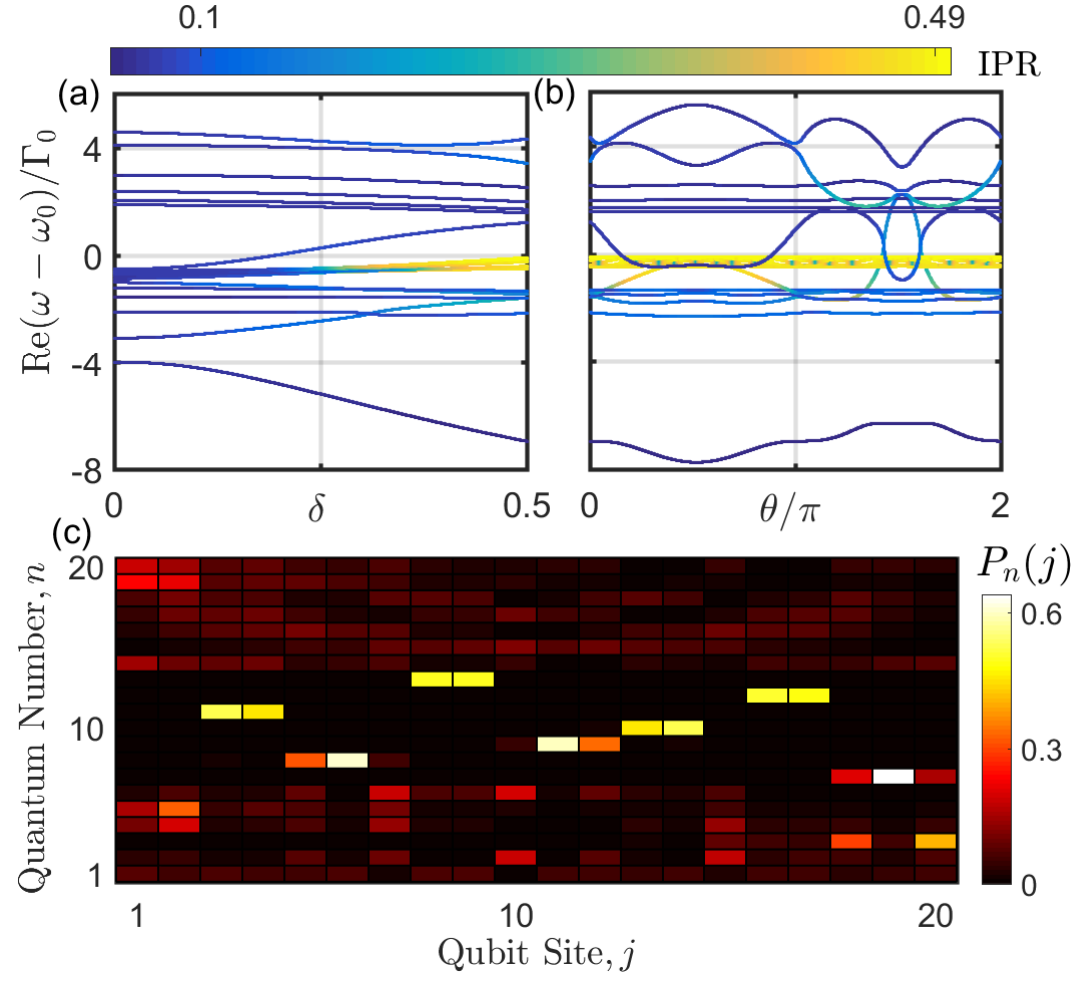}
	\caption{Energy spectrum as a function of (a) quasiperiodic strength by fixing $\theta=0$ and (b) modulation phase by fixing $\delta=0.5$. The colors denote the inverse participation ratio ($\textrm{IPR}$) of the corresponding eigenstates.
		(c) Probability distribution in the qubit array with modulation strength $\delta=0.5$ and modulation phase $\theta=0$. The colors denote the probability $P_n(j)$. 
		The other parameters are chosen as $\omega_0=100$, $\beta=(\sqrt{5}+1)/2$, $\Gamma_0=0.01$, and $\varphi=\omega_0 d/c=1$.
	}\label{fig:1}
\end{figure}
%%%%%%%%%%%%%%%%%%%%%%%%%%%%%
%
The eigenstates and eigenvalues are obtained by diagonalizing $H_{eff}|\psi_n\rangle=\omega_n|\psi_n\rangle$ in the single-excitation subspace. 
To characterize the localization of excitation, we calculate the inverse participation ratio (IPR)
\begin{equation}
	\textrm{IPR}=\sum\limits_j |\psi_n(j)|^4,
\end{equation}
where $\psi_n(j)$ is the amplitude of the excitation state at the $j$th site.
For $\textrm{IPR}\rightarrow 1$, the excitation is completely located at a single qubit.
For $\textrm{IPR}\rightarrow 0$, the excitation is completely delocalized throughout the qubit array.
Thus, the larger $\textrm{IPR}$ indicates that a single excitation is more localized.

We show the energy spectrum as a function of quasiperiodic strength; see Fig.~\ref{fig:1}(a). 
The parameters are chosen as $\omega_0=100$, $\beta=(\sqrt{5}+1)/2$, $\Gamma_0=0.01$, and $\varphi=\omega_0 d/c=1$.
The colors in the spectrum indicate $\textrm{IPR}$ of the corresponding eigenstates.
As the quasiperiodic strength increases, some eigenstates around the resonant frequency $\omega_0$ change from delocalization to localization.
However, most of the eigenstates away from the resonant frequency still keep delocalized, except for several edge states.
For a fixed large quasiperiodic strength, we can find that there exist delocalization-to-localization transitions as the energy increases. 
To make it clear, we calculate the energy spectrum as a function of the modulation phase; see Fig.~\ref{fig:1}(b). 
The other parameters are chosen the same as those in Fig.~\ref{fig:1}(a) except for $\delta=0.5$.
The localized states form a continuum band around the resonant frequency.
As the modulation phase changes, the delocalized states may penetrate into the continuum band of localized states.
However, although the penetrated delocalized states have an energy close to that of the localized states, they do not couple with the localized states.
The maximum of $\textrm{IPR}$ can be up to $1/2$, a feature value for an equal population of an excitation in two qubits.
Indeed, there are several states that are mostly populated in two neighboring qubits; see the probability distribution of all the eigenstates in Fig.~\ref{fig:1}(c).
The parameters are chosen the same as those in Fig.~\ref{fig:1}(a)  except for $\delta=0.5$.
The $y$ axis is the quantum number of eigenstates arranged in the order of incremental energy. 
The localized states in the continuum band are mainly located in the bulk, apart from several localized edge states.

In stark contrast to Ref.~\cite{fayard2021many}, in which all the eigenstates are localized in a completely disordered qubit array as the system size tends to be infinite,
the localized states in the continuum band are only a fraction of the total eigenstates.
In the case of $(\sqrt{5}+1)/2$, the fraction of localization is up to $(3-\sqrt{5})/2$.

%%%%%%%%%%%%%%%%%%%%%%%%%%%%%
\begin{figure}[!htp]
	\includegraphics[width=0.52\textwidth]{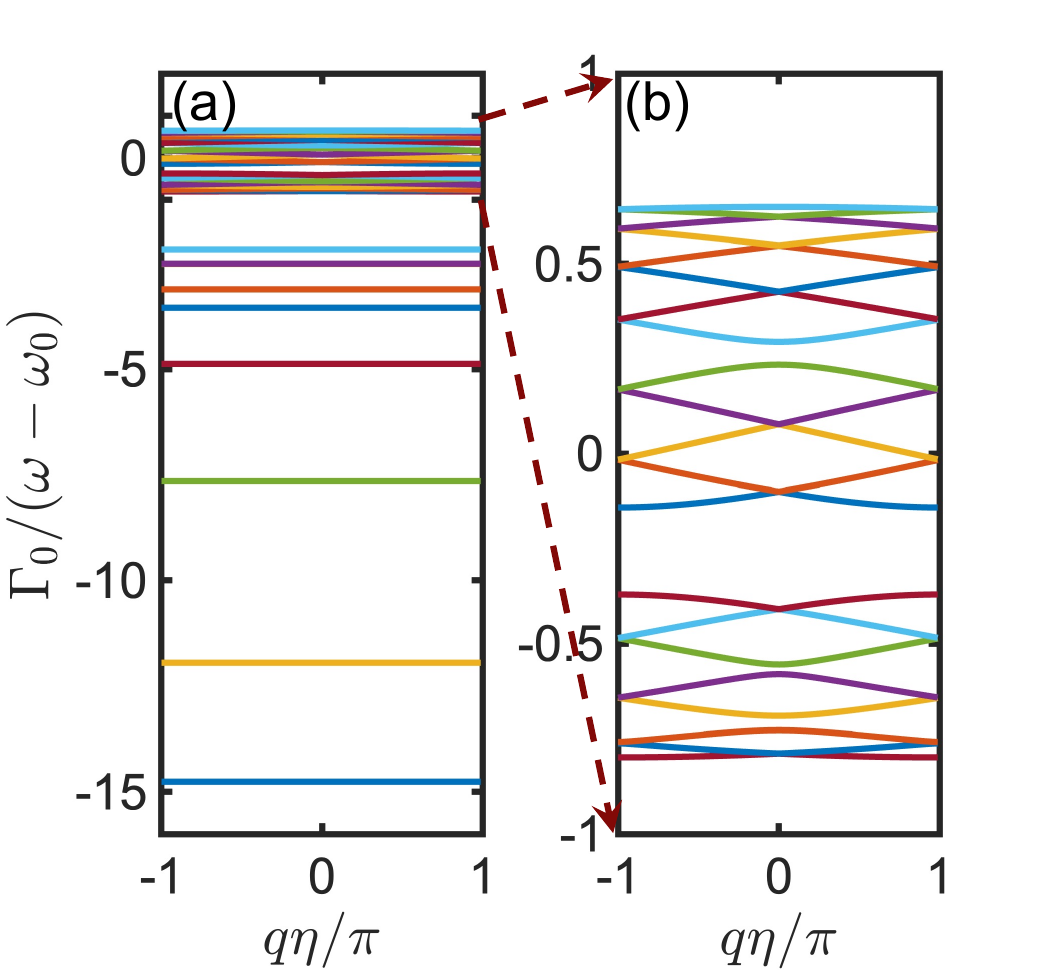}
	\caption{(a) Inverse energy band in the large periodic approximation $F_{n+1}/F_n=55/34$. (b) is the enlarged view of (a) around $\Gamma_0 S_q\in [-1,1]$.  The other parameters are chosen as $\omega_0=100$, $\Gamma_0=0.01$, $\theta=0$, and $\varphi=\omega_0 d/c=1$.
	}\label{fig:5}
\end{figure}
%%%%%%%%%%%%%%%%%%%%%%%%%%%%%

\subsection{Fraction of localization in the periodic approximation} \label{Sec5}
To derive the localization fraction, we consider the periodic approximation of $\beta=(\sqrt{5}+1)/2$ by using a fraction representation, $\chi/\eta =(1/1,2/1,3/2,5/3,8/5,...,F_{n+1}/F_{n},...)$ with the Fibonacci sequence $F_n$, which satisfies $F_n=F_{n-1}+F_{n-2}$.
For the rational ratio with larger $F_n$, the quasiperiodic WQED makes a transition to a periodic system with translational symmetry, but still maintains some properties of the spatial modulation.
In such a large-period limit, we can obtain the band structure.

According to the Bloch Theorem, the eigenstates are Bloch states. 
\begin{equation}
	|{\psi _q}\rangle  = \sum\limits_{j,l}^{} {{e^{iq\eta j}}{u _l}|\eta j + l\rangle }, \label{Bloch}
\end{equation} 
where $q\in [-\pi/\eta,+\pi/\eta]$ is a quasimomentum of an excitation, $j$ is the cell index, $l$ denotes the sublattice in a cell, and $u_l$ are periodic functions of the Bloch states.
Substituting Eq.~\eqref{Bloch} into  the Schr\"odinger equation $H_{eff}|\psi_q\rangle =\omega_q |\psi_q\rangle$, the equations for $\{u_l\}$ are given by
\begin{eqnarray}
	({\omega_q} - {\omega _0})/\Gamma_0{u_{l'}} = \sum\limits_{l = 1}^\eta  {{H_q}(l,l'){u_l}},
\end{eqnarray} 
with elements of the Bloch Hamiltonian 
\begin{eqnarray}
	{H_q}(l,l') &=&\sin \left({\varphi |{z_{l'}} - {z_l}|} \right) + \frac{{i\sin(q\eta )\sin \left[ {\varphi ({z_{l'}} - {z_l})} \right]}}{{\cos(q\eta ) - \cos(\varphi \eta )}} \nonumber \\
	&+& \frac{{\sin (\varphi \eta )}\cos \left[ {\varphi ({z_{l'}} - {z_l})} \right]}{{\cos(q\eta ) - \cos(\varphi \eta )}},
\end{eqnarray}
where $H_q$ becomes a Hermitian matrix $H_q=H_q^\dagger$. It means that
the energy band of the excitation becomes real, because there is no way for the excitation to decay when the qubit number tends to be infinite.

To avoid the divergence of the energy band at $q=\pm \varphi$,
we define inverse energy band as~\cite{PhysRevLett.131.103604}

\begin{equation}
	S_q=\frac{1}{\omega_q-\omega_0},
\end{equation}
which can be obtained by diagonalizing the inverse of the Bloch Hamiltonian $H_q$.
We first calculate the inverse energy band $S_q$ as a function of the quasi-momentum; see Fig.~\ref{fig:5}(a).
The inverse energy band can be classified as flat bands in the region $\Gamma_0|S_q|\ge 1$, and curved bands in the region $\Gamma_0 | S_q| \le 1$.
The flat inverse bands mean that the group velocities of these excitation states are extremely small.
Even a tiny residue quasiperiodic modulation can make the slow excitation states stuck and localized.
Indeed, the localized states around the resonant frequency have $|\omega-\omega_0|\ll \Gamma_0$ and coincide with flat inverse bands. 
The flat inverse bands contribute to the localization of excitation.
To better view the curved inverse bands, we enlarge the view around $|\Gamma_0/(\omega-\omega_0)|\le 1$ in Fig.~\ref{fig:5}(b).
Because of the large dispersion of curved inverse bands, the group velocity of an excitation is so large that a tiny residue quasiperiodic modulation cannot localize such an excitation.
Thus, the curved inverse bands contribute to the extended wave of excitation.
The localization fraction is equal to the fraction of flat inverse band.
The numbers of flat inverse bands and curved inverse bands are $F_{n-2}$ and $F_{n-1}$, respectively.
Thus, the fraction of localization is up to $F_{n-2}/F_n=(3-\sqrt{5})/2$ as $F_n$ tends to be infinite.

\subsection{Scattering of a single photon} \label{Sec4}
%%%%%%%%%%%%%%%%%%%%%%%%%%%%%
\begin{figure}[!htp]
	\includegraphics[width=0.48\textwidth]{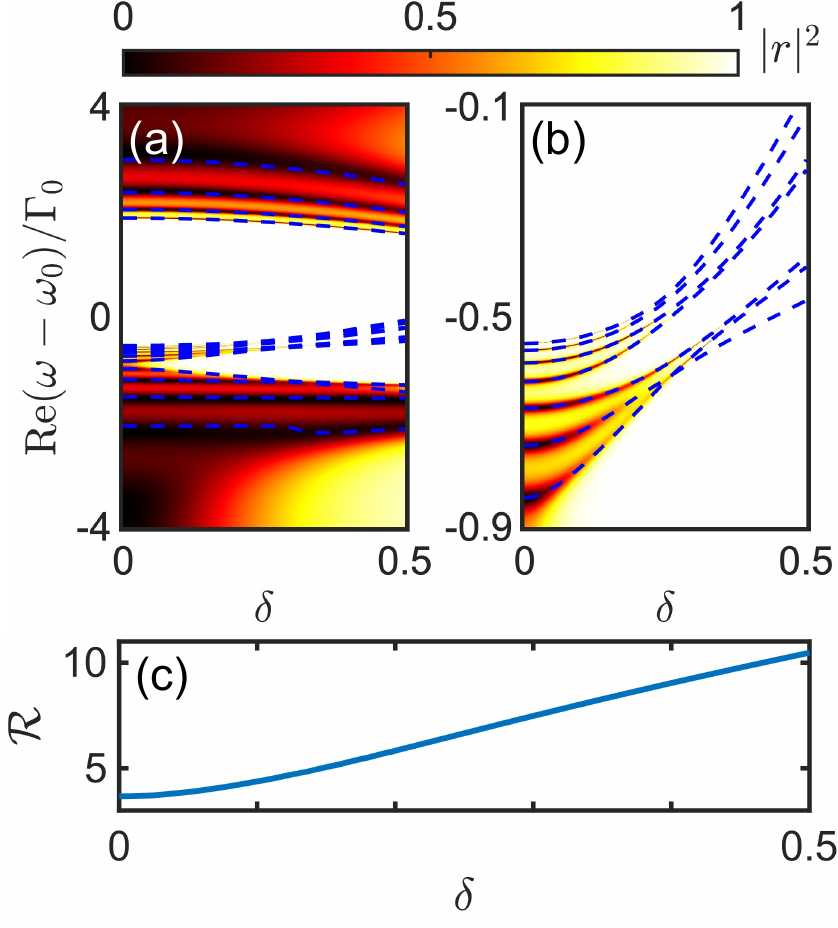}
	\caption{(a) Reflection of a single photon as a function of quasiperiodic strength and frequency of photon, and (b) its enlarged view.
		 (c) The overall reflection as a function of quasiperiodic strength.
		The parameters are chosen as $\omega_0=100$, $\beta=(\sqrt{5}+1)/2$, $\Gamma_0=0.01$, $\theta=0$, $\varphi=\omega_0 d/c=1$.
	}\label{fig:3}
\end{figure}
%%%%%%%%%%%%%%%%%%%%%%%%%%%%%

%%%%%%%%%%%%%%%%%%%%%%%%%%%%%
\begin{figure}[!htp]
	\includegraphics[width=0.49\textwidth]{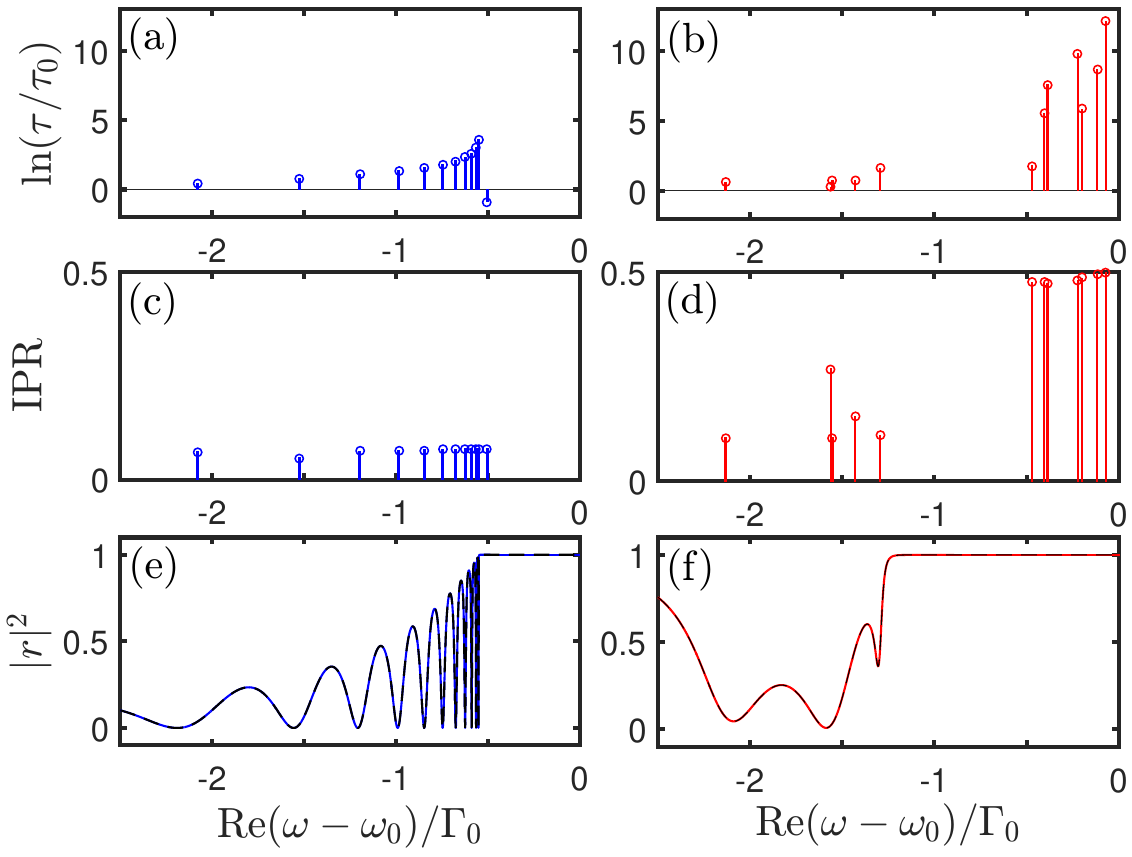}
	\caption{Comparison between properties of excitation states and the reflection of photon. The left and right panels are the cases with no quasiperiodic modulation ($\delta=0$) and strong quasiperiodic modulation ($\delta=0.5$).
		(a) and (b) give the lifetime of eigenstates. (c) and (d) give the $\textrm{IPR}$ of eigenstates.
		(e) and (f)  show reflection as a function of frequency of a single photon. The blue and red solid lines are calculated via Green function method in momentum space, and the black-dashed line is calculated via transfer-matrix method in real space. The other parameters are chosen as $\omega_0=100$, $\Gamma_0=0.01$, $\theta=0$, $\beta=(\sqrt{5}+1)/2$, and $\varphi=\omega_0 d/c=1$.
	}\label{fig:4}
\end{figure}
%%%%%%%%%%%%%%%%%%%%%%%%%%%%%

We are curious about how the reflection or transmission of a single photon is affected by the quasiperiodic modulation of the qubit spacings. 
We calculate the reflection as a function of the quasiperiodic strength and frequency of an injected photon; see Fig.~\ref{fig:3}(a) and Fig.~\ref{fig:3}(b) for its enlarged view near $\omega_0$.
Because the subradiant states strongly affect the reflection, we also show the spectrum of subradiant states as a function of quasiperiodic strength; see the dashed blue lines. 
The parameters are chosen as $\omega_0=100$, $\Gamma_0=0.01$, $\delta=0.5$, $\beta=(\sqrt{5}+1)/2$, $\theta=0$, and $\varphi=\omega_0 d/c=1$.
The dip of the reflection appears when the photon is in resonant with the subradiant states, except for those near resonant frequency at the large quasiperiodic strength.
We find that those exceptional subradiant states are localized states which have large $\textrm{IPRs}$, as shown in Fig.~\ref{fig:1}. 
It seems that the delocalized (localized)  subradiant states suppress (enhance) the reflection of a single photon.    
To estimate the effect of quasiperiodicity on reflection, we define an overall reflection as
\begin{equation}
	\mathcal {R}=\frac{1}{\Gamma_0}\int_{0}^{+\infty} |r_{\omega/c}|^2d\omega.
\end{equation} 
where the integral is taken over the frequency of an injected photon.
When $\omega$ tends to $0$ and $+\infty$, the reflection approach to $0$. Actually, only the photon with frequency around the resonant frequency $\omega_0$ plays a significant role. 
The overall reflection monotonically increases with the quasiperiodic strength; see Fig.~\ref{fig:3}(c).

To be clear, we calculate the lifetime and $\textrm{IPR}$ of the excitation eigenstates, and compare them to the reflection as the frequency of photon changes; see Fig.~\ref{fig:4}.
The left and right pannels are obtained without quasiperiodic modulation ($\delta=0$) and with strong quasiperiodic modulation ($\delta=0.5$), respectively. 
Here, we only show the results below the resonant frequency to give a better vision.
The lifetime is given by $\tau=1/\textrm{Im}(\omega_n)$, which can be used to distinguish subradiant states from superradiant states.
Subradiant states have a lifetime $\tau/\tau_0>1$ and superradiant states have a lifetime $\tau/\tau_0<1$, where $\tau_0=1/\Gamma_0$ is the lifetime of a single excitation.
In the case without quasiperiodic modulation, most of the excitation states are subradiant states below resonant frequency $\omega_0$, except for one superradiant state around $\omega_0-\Gamma_0/2$; see Fig.~\ref{fig:4}(a).
All eigenstates are delocalized states and have small $\textrm{IPR}$; see Fig.~\ref{fig:4}(c).
We find that the reflection spectrum is at a dip when the frequency of the photon is in resonance with a delocalized subradiant state; see Fig.~\ref{fig:4}(e).
It means that the delocalized subradiant states suppress the reflection of the photon and reversely enhance the transmission of the photon.  
Moreover, the width of the dip is also related to the lifetime of the delocalized subradiant states.
Generally, if the delocalized subradiant states have a longer lifetime, the resonant photon is reflected in a narrower frequency range. 
The photon is completely reflected when its frequency lies in the bandgap, consistent with previous results~\cite{Brehm2021o}.

In the case with strong quasiperiodic modulation ($\delta=0.5$), the eigenstates shown in the energy window are subradiant states; see the lifetime in Fig.~\ref{fig:4}(b).
The lifetime is dramatically increased for the subradiant localized states with larger $\textrm{IPR}$.  
This is mainly due to the fact that the localized subradiant states are strongly confined in the bulk, away from the edge qubits which are the photon leakage channel for losing excitation~\cite{Poddubny2020}.
We find that the photon is almost perfectly reflected when its frequency is in resonance with the localized subradiant states.
Similarly to the first case, there are delocalized subradiant states in the dip of the reflection spectrum.
Thus, we have further confirmed that the localized (delocalized) subradiant states block (enhance) the transmission of photon.

In Figs.~\ref{fig:4}(e) and \ref{fig:4}(f), we have to emphasize that the reflection of a single photon is calculated via the transfer-matrix method in real space (red  or blue solid lines) and Green function method in momentum space (black dashed lines).
The exact same results indicate that the two methods are equivalent.

\subsection{Transmission mobility edge} 
\label{Sec6}
%%%%%%%%%%%%%%%%%%%%%%%%%%%%%
\begin{figure}[!htp]
	\includegraphics[width=0.51\textwidth]{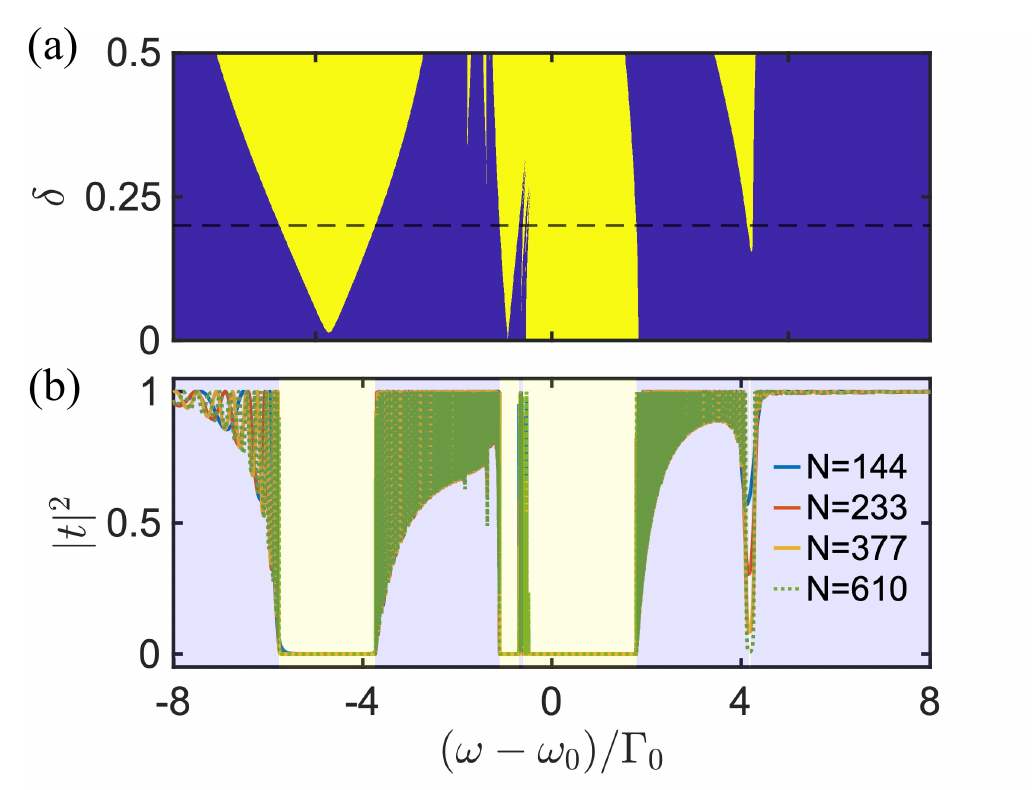}
	\caption{(a) Localization and delocalization phases in transmission. The yellow (blue) region corresponds to the localized (delocalized) phase. Here, the transfer-matrix method is used to calculate the resistance, and the maximum number of atoms reaches $N=2584$. (b) shows transmission as a function of photonic frequency for different system sizes. The blue, red, yellow solid lines and green dotted line correspond to atom numbers $N=144, 233, 377, 610$, respectively. The modulation strength is fixed at $\delta$=0.2, corresponding to the black dashed line in panel (a). Other parameters are $\omega_0=100$, $\Gamma_0=0.01$, $\theta=0$, $\beta=(\sqrt{5}+1)/2$, and $\varphi=\omega_0 d/c=1$.
	}\label{fig:6}
\end{figure}
%%%%%%%%%%%%%%%%%%%%%%%%%%%%%
%
Anderson localization arises from destructive interference induced by static disorder in a system \cite{PhysRev.109.1492}.
Although in one- and two-dimensional Anderson systems, all states are typically localized.
But in the three-dimensional systems, there exists the mobility edge that separates localized from delocalized energy eigenstates \cite{Semeghini2015}.
In contrast to random disorder, quasiperiodic systems can exhibit a mobility edge even in one spatial dimension.
For instance, the mobility edge emerges in generalized Aubry-André-Harper (AAH) models with on-site modulations or long-range couplings \cite{doi:10.1126/science.aaa7432,PhysRevLett.114.146601,PhysRevLett.123.025301,PhysRevLett.126.040603}.
We also observe the mobility edge in quasiperiodic waveguide QED system, which manifests as a sharp change in transmission at the critical frequency of a single photon.

As is well known, the signature of Anderson localization is the typical resistance grows exponentially with the length of system \cite{PhysRevB.24.5583,JLPichard_1986}.
We calculate the resistance in real space using the transfer-matrix method.
This allows us to identify the mobility edge, at which the excitation eigenstates undergo localization transition.
The dimensionless resistance is given by the Landauer formula \cite{Landauer01041970,PhysRevB.24.2978}
\begin{equation}
	\rho=\frac{1-|t_\kappa|^2}{|t_\kappa|^2},
\end{equation}
where $t_\kappa$ is the transmission coefficient.
Using the property $T_{11} T_{22} - T_{12} T_{21} = 1$ of transfer matrix, the resistance can be expressed as $\rho=T_{12} T_{21}$.

According to the method in Refs.~\cite{PhysRevB.24.5583,JLPichard_1986}, if the system is localized, the resistance is exponential increased with the system size, otherwise it is delocalized. 
We calculate the resistance as a function of system size $N$ with different quasiperiodic strength and photonic frequency. 
If the scaling of resistance obeys feature of localization/delocalization, we color them as yellow/blue in the parameter space, respectively. 
Fig.~\ref{fig:6}(a) shows a color map in the parameter space.
Our numerical analysis extends to a maximum system size of $N = 2584$, a Fibonacci number chosen to provide an accurate rational approximation of the quasiperiodic modulation.
From this phase diagram, we can find sharp change from localization phase to delocalization phase as photonic frequency varies.
In analog to the conventional definition of mobility edge, we term the localization-delocalization phase boundary in photonic frequency as transmission mobility edge.  
The transmission mobility edge separate the localization phase that blocks transmission of photons and delocalization phase that support finite transmission of photons.

There are three major localization regions in the parameter space.
The localization appears only for finite modulation strength $\delta$, in the left localization region around $(\omega-\omega_0)/\Gamma_0\sim -4.7$ and the right localization region around $(\omega-\omega_0)/\Gamma_0\sim 4$.
These localization can be explained by the introduction of quasiperiodic modulation.
However, localization exists even in the absence of quasiperiodic modulation in the middle region around $\omega_0$.
This can be explained by large band gap of excitation around resonant frequency.
However, as quasiperiodic strength increases, there also exist localized excitation states around the band gap.
Hence, we can deduce that the localization is a joint effect of band gap and localized excitation states for larger quasiperiodic strength.

To further verify the validation of mobility edge, we fix  quasiperiodic strength $\delta=0.2$ and calculate the transmission as a function of photon frequency for different system sizes.
To obtain optimal rational approximation of $\beta=(\sqrt{5}+1)/2$, we set the number of atoms to a sequence of Fibonacci numbers, $N=144$, $233$, $377$, and $610$.
%When the frequency lies between the pair of mobility edges,
The transmission will decrease with system size in the localization phase and maintain finite values in the delocalization phase.
Comparing Figs.~\ref{fig:6} (a) and (b), one can find that in the light yellow regions, the transmission indeed decreases as $N$ increases, corresponding to the localization region.
However, in the light blue regions, the transmission oscillates around finite values as $N$ increases, corresponding to the delocalization region.
The transmission mobility edges can be exactly obtained from the analysis of finite-size scaling.

\section{Summary and discussion}
We have studied the localization and scattering of a photon in qubit arrays with quasiperiodic modulation of spacing.
We have developed transfer-matrix method in real space and Green function method in momentum space to calculate the scattering of photon, and showed that the two methods give exactly the same results.
As the strength of quasiperiodic modulation increases, the subradiant states gradually change from extended states to localized states.
The strong quasiperiodic modulation leads to the localization of fractional eigenstates near resonant frequency, in stark contrast to disordered WQED systems where all the eigenstates are localized in the thermodynamic limit.
The localized and delocalized states are respectively from the flat inverse bands and curved inverse bands in the limit of large-period modulation.
The single localized excitation blocks the transmission of a photon in resonance with the excitation, but the delocalized excitation can support the transmission. 
Furthermore, we also find transmission mobility edge, where delocalization-localization transition in transmission happens as photonic frequency changes.

In this work, we only consider how localized single-excitation subradiant states affect the scattering of a single photon.
When considering more photons, many-body localization of excitations is predicted in disordered WQED systems when the number of photons is less than half of the chain~\cite{fayard2021many}.
The many-body localization of multi-excitations may play a significant role in the inelastic scattering of multiple photons.
Compared to disordered WQED systems, the quasiperiodic WQED systems may have richer novel multi-excitation states, such as products of delocalized states and localized states.
It is deserved to develop a more powerful method to systematically study the role of multi-excitation localization states in inelastic scattering of multiple photons and the transmission mobility edge.

\begin{acknowledgements}
We thank Ling Lin and Na Zhang for helpful discussions. This work is supported by the National Natural Science Foundation of China (Grants No. 12275365, 12025509, and 92476201), the National Key Research and Development Program of China (Grant No. 2022YFA1404104), the Guangdong Provincial Quantum Science Strategic Initiative (GDZX2204003, GDZX2305006 and GDZX2405002), and the Natural Science Foundation of Guangdong Province (Grant No. 2023A1515012099).
\end{acknowledgements}

\nocite{*}
%apsrev4-2.bst 2019-01-14 (MD) hand-edited version of apsrev4-1.bst
%Control: key (0)
%Control: author (72) initials jnrlst
%Control: editor formatted (1) identically to author
%Control: production of article title (-1) disabled
%Control: page (1) range
%Control: year (1) truncated
%Control: production of eprint (0) enabled
%

%\bibliography{QuasiperiodicWQED}

\end{document}